# Features of coal dust dynamics at action of differently oriented forces in granular filtering medium

I. M. Neklyudov, L. I. Fedorova, P. Ya. Poltinin and O. P. Ledenyov

*National Scientific Centre Kharkov Institute of Physics and Technology, Academicheskaya 1, Kharkov 61108, Ukraine.*

The process of the coal dust particles transportation and structurization in the experimental horizontal model of air filter with the cylindrical coal adsorbent granules as in the iodine air filter *AU-1500* at the nuclear power plant is researched. In the investigated case the vector of carrying away force of air flow and the vector of gravitation force are mutually perpendicular, and the scattering of the dust particles on the granules occurs in the normal directions. It is found that the phenomenon of non-controlled spontaneous sharp increase of aerodynamic resistance in the iodine air filter under the big integral volumes of filtered air and the big masses of introduced coal dust particles is not observed at the described experimental conditions in distinction from the case of the parallel orientation of this forces as in the vertical iodine air filters at the nuclear power plant. The quantitative measurements of the main parameters of the process of the dust masses transportation and structurization are made on a developed experimental model of the iodine air filter with the cylindrical coal adsorbent granules.

## Introduction

It is known that, after a long term period operation at the nuclear power plants (*NPP*), the sharp abnormal increase of aerodynamic resistance is observed in the iodine air filters (*IAF*) of the model *AU-1500* with the cylindrical coal granules filtering medium (*GFM*). In [1], it was found that the *AU-1500* adsorbers failures are usually connected with the appearance of dispersive coal dust particles fraction as a result of the destruction of cylindrical coal granules at the absorber's input surface during the air stream flow process in the vertically positioned *IAF* at the *NPP*. The physical features of processes of subsequent transportation and structurization of small coal dust particles, which are accompanied by the creation of the air stream flow blocking layer, were more comprehensively researched in [2]. This investigations on the distribution of small dispersive coal dust particles along the absorber in the case, when the air-dust aerosol was supplied in small parts into the vertically positioned adsorber, is completed, using the model of the *AU-1500* filter. It was shown that the distribution of small coal dust particles along the filter has a gradient dependence so that there are a big maximum of the particles accumulation in the sub-surface layer of filter input and a number of small maximums of the accumulation insight the absorber in the *IAF*. The parallel analysis of both the obtained experimental results on the *IAF* models parameters and the study of the adsorbed fractions contents in the failed *IAFs* allowed us to find a root cause of the sharp increase of aerodynamic resistance in the *IAF* after its long term period operation at the *NPP*. The necessary recommendations on the possible improvement of the *IAF* design, aiming to increase the *IAF* operational cycle, were also provided.

Going from the analysis in [2], the scope of our research interest is focused on the investigation of the *IAF* with the changed space orientation in relation to the direction of physical forces, acting on the dust masses. Therefore, we conduct the research on the adsorber, placed in the horizontal position. In this case, we expect to observe the decrease of influence by the transported small dispersive coal dust masses on the aerodynamic resistance of the *IAF* due to the change of the physical conditions of experiment.

Let us note that the small dispersive coal dust particles with the discrete numbers of dimensions have been created during the seizing of the adsorbent of the type of *CKT-3* in the *AU-1500* adsorber in the *IAF* [2, 3].

At the vertical orientation of the *IAF*, the physical forces, which are directed toward the one direction (from the top to the bottom), mainly act on the small dispersed coal dust particles, which are transported by the air-dust aerosol in the *IAF*. These physical forces include the capturing force, defined by the *Stocks* force $F_S$, and the gravitation force $F_g$. At an action by these physical forces, the small dispersive coal dust particles fraction is being transported by the air – dust aerosol in the *IAF*. The average velocity of transportation of small dispersive coal dust particles is defined by their geometrical dimensions. The smallest coal dust particles reach the highest velocities, accelerating by the viscous capturing force, which is a main moving force for the smallest particles during their transportation by the air-dust aerosol inside the *IAF*. The biggest coal dust particles have the lowest velocity and the dynamics of their movement inside the *IAF* is strongly influenced by the gravitation force $F_g$. They are weakly captured by the air-dust aerosol, and found to be the first coal dust particles, which usually precipitate between the cylindrical coal granules at the top part of absorbent layer, resulting in the creation of the equally distributed dense coal dust structure with the coal dust particles of all the dimensions at the sub-surface layer of absorber's input during a long term period of absorber's operation at the *NPP* [4]. The reaching of critical value of mass share of small dispersive coal dust particles in the blocking layer results in an exponential increase of aerodynamic resistance of the *IAF*. It was interesting to find out the answer on the following question:

*1)* How will the small dispersive coal dust particles fraction, appearing in the close proximity to the surface of granular filtering medium at the absorber's input, be distributed along the horizontally oriented *IAF*, when the physical forces, acting on the small dispersive coal dust particles are perpendicular?

In this case, the viscous capturing force transports the small dispersive coal dust particles in the granular filtering medium along the absorber, whereas the gravitation force $F_g$ moves the small disperse coal dust particles in the granular filtering medium toward the bottom of absorber in the horizontally oriented *IAF*. Also, in this experimental approach, it will be necessary to clarify the following questions:

*1)* How will the small dispersive coal dust particles be distributed at the cross-section of the absorber in the horizontally oriented *IAF*?

*2)* What is the exact quantity of introduced small dispersive coal dust particles, which will be jettisoned outside the *IAF* at the air filtering process in the horizontally oriented *IAF*?

*3)* What is the shape of dependence between the magnitude of aerodynamic resistance and the physical features of small dispersive coal dust particles distribution in the granular filtering medium of absorber in the horizontally oriented *IAF*?

The experimental research with the precise quantitative measurements, aiming to make the accurate characterization of main parameters of small dispersive coal dust particles transportation process in the granular filtering medium with the cylindrical coal adsorbent granules in the horizontally oriented *IAF*, are completed, using the horizontally oriented developed model of the *IAF*, which was exposed to the physical processes of the small dispersive coal dust particles generation, structurization and transportation.

## Experimental Measurements Procedure

In Fig. 1 (a), the developed model of the *IAF* with the horizontally oriented glass cylinder with the granular filtering medium, is shown. The developed model of the *IAF* is in the *10* times smaller than the diameter of adsorber in the real *IAF* at *NPP*.

a)

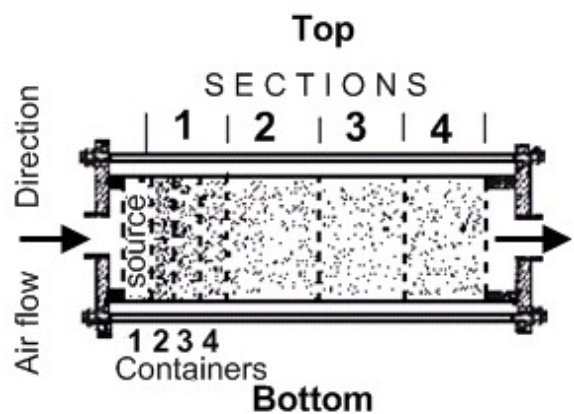

b)

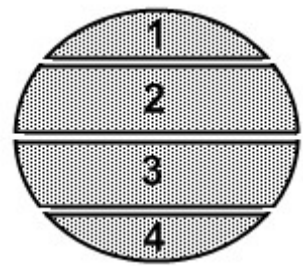

*Fig. 1. Scheme of iodine air filter model (IAF):*
*a – Model of adsorber;*
*b – Transverse cross-section of container;*
*1-4 – Segments, from top to bottom.*

In our experiment, the absorbent layer was fixed between the two nets in the model of the *IAF*. The air stream flow was directed from the left position to the right position and the filtered volume of air was in the *100* times lower than in the case of the real operational *IAF* at the *NPP*. However, the average velocity of air stream flow $(0.1 - 0.58)$ $m/sec$ was approximately equal to the average velocity of air stream flow during the standard air filtering process in the *AU-1500*. The cylindrical coal granules of the type of *CKT-3* was used and the length of adsorbent layer in the *IAF* model was comparable to the height of adsorbent layer in the *AU-1500*. The magnitude of aerodynamic resistance of the developed *IAF* model was comparable with the aerodynamic resistance in the real *IAF* at the *NPP*.

The layer with the cylindrical coal granules was divided by the special nets on the four sections $(1 - 4)$, calculated from the left side position to the right side position). The section no. *1* includes the three metallic containers no. *2, 3, 4* designed so that every container has the layer of cylindrical coal granules, fixed by the special nets (the height of layer is *2 cm*). The granular filtering medium with the cylindrical coal granules in each containers was divided by the special nets on the four segments (*1 − 4*) in Fig. 1 (b)) with the purpose to make it possible to evaluate the physical character of distribution of small dispersive coal dust particles at the transverse cross-section of every container. The container no. *1*, representing an experimental source of small disperse coal dust particles, is attached to the container no. *2* in the filter model.

In the experiments, the adsorber (part with the granular filtering medium) contained the homogenous cylindrical coal adsorbent granules with the length of *3.2 mm* and the diameter of *1.8 mm* of the type of *СКТ-3*. The small dispersive coal dust particles with the dimensions of *10 μm* were produced by the rubbing of the adsorbent *СКТ-3* granules during the special process before their subsequent loading to the container no. *1* in the filter model. Before the beginning of every new experiment, the newly prepared mixture between the cylindrical coal granules and the small dispersive coal dust particles was loaded to the container no. *1* (a source of dust) with the purpose to keep the same starting experimental conditions. Thus, the accumulation of small dispersive coal dust particles in the container no. *1* was fully excepted. The mass of small dispersive coal dust particles did not exceed the value of *1,5 %*, comparing to the total mass of cylindrical coal granules in the source. It is necessary to explain that the small dispersive coal dust was introduced by the discrete parts to the model of the *IAF*.

The mass of every container from the four containers was precisely measured before the beginning of new experiment and after its completion. The total mass of introduced small dispersive coal dust in the model of the *IAF* was calculated, measuring the initial mass and the final mass of the container no. *1*. In the cases of the containers no. *2–4*, the mass of the precipitated small dispersive coal dust particles in every container in the model of the *IAF* was determined as a difference between the initial mass of container and the final mass. The mass of coal dust, jettisoned outside the containers into the model, was computed as a difference between the total dust mass introduced into the model and the total seized dust mass in the sections no. *2–4*.

The following technical designations are used:
$M_0$ is the mass of cylindrical coal granules in all the *IAF*;
$M_2, M_3, M_4$ are the masses of granules in the containers no. *2, 3, 4* correspondingly;
$m_0$ is the total mass of small dispersive coal dust particles, introduced into the *IAF* after the completion of experiment;
$m_2, m_3, m_4$ are the total masses of small dispersive coal dust particles, precipitated in the containers no. *2, 3, 4* after the completion of experiment;
$m_4$ is the total mass of small dispersive coal dust particles, jettisoned outside the sections no. *2–4* after the completion of the experiment, ($m_4 = m_0 − m_1 − m_2 − m_3$);
$M_i^j$ is the mass of adsorber in the segment *j* (*j* changes from *1* to *4*) of the container no. *i* (*i* changes from *2* to *4*);
$m_i^j$ is the mass of small dispersive coal dust particles, accumulated in the segment *j* in the container no. *i*;

$m_{0i}$ is the mass of small dispersive coal dust particles, precipitated in the container no. *i* at the end of experiment.

## Measurements results on spatial distribution of coal dust particles in granular filtering medium in iodine air filter

The research results on the dependence of the relative mass of the small dispersive coal dust particles fraction, accumulated in the containers no. *1–3*, $m_1/m_0$, $m_2/m_0$, $m_3/m_0$, (the graphs: *1–3*), and the relative mass of the small disperse coal dust particles fraction, disappeared in the containers no. *1–3*, $m_4/m_0$, (the graph *4*) on the mass share of the small disperse coal dust particles fraction, introduced in the *IAF*, $m_o/(M_o + m_o)$, are shown in Fig. 2.

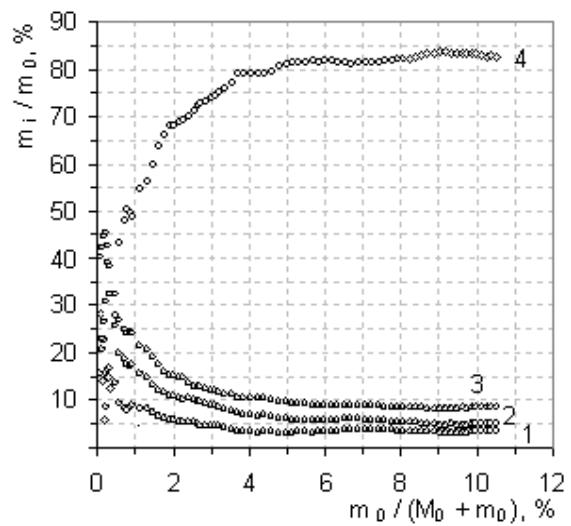

*Fig. 2. Dependence of relative mass of small disperse coal dust, precipitated in first container (1), second container (2), third container (3) and relative mass of small disperse coal dust, disappeared in sections 1–3 (4) on mass share of small disperse coal dust fraction, appeared in iodine air filter; i changes from 1 to 4.*

The research results on the dependence of the small disperse coal dust particles fraction, precipitated in the containers no. *1–3* at the levels of $m_1/(M_1 + m_1)$, $m_2/(M_2 + m_2)$, $m_3/(M_3 + m_3)$, on the mass share of the small disperse coal dust fraction, introduced in the *IAF*, $m_o/(M_o + m_o)$, are shown in the graphs *1–3* in Fig. 3.

The comparative analysis of obtained experimental data in Figs. 2 and 3 allows us to make the following conclusions:

**1**. In the horizontally oriented *IAF*, the chaotic re-distribution of the small dispersive coal dust particles at the height of adsorbent layer is observed during the introduction of the small particles in the form of small discrete parts in the filter at the initial observation time-period [up to *0.5 %* at $m_o/(M_o + m_o)$]. The clearly observable trend toward the decrease of quantity of the coal dust particles, precipitated in the containers no. *1–3*, is registered at the specific experimental parameters: from *0.5%* to *5%* at the axis, $m_o/(M_o+m_o)$, as shown in the graphs *1–3* in Fig. 2.

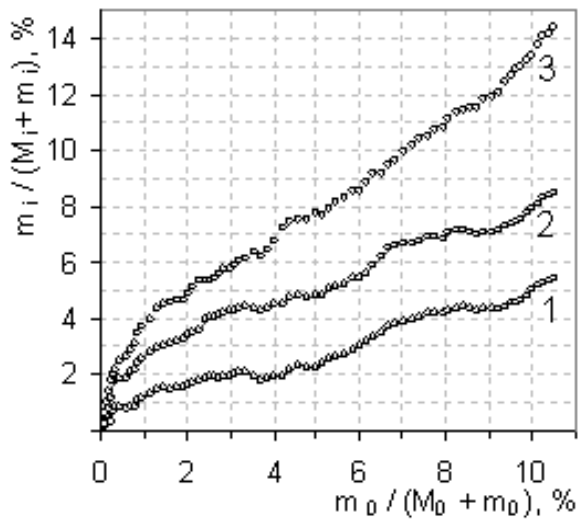

*Fig. 3. Dependence of relative mass of small disperse coal dust, precipitated in first container (1), second container (2), third container (3) on mass share of small disperse coal dust fraction , appeared in iodine air filter (IAF); i changes from 1 to 3.*

**2**. Moreover, the above denoted physical trend toward the decrease of quantity of the small dispersive coal dust particles, precipitated in the containers no. *1–3*, is sharply expressed at the magnitudes from *0,5 %* to *3%* (the decrease on the values of *9%; 16%* and *21%* in the cases of containers no. *1, 2* and *3* correspondingly). The relative mass share of the dust particles, precipitated in the containers no. *1–3*, is increased at the levels of *1%; 2.4%* and *3.4%* (see the graphs *1–3* in Fig. 3). The mass share of the dust particles in the *3$^{rd}$* container, which was situated near the source of dust, is in the *1.4* times bigger than the mass share of the dust particles in the *2$^{nd}$* container; and it is in the *3* times bigger than the mass share of the dust particles in the *1$^{st}$* container. (The mass share of the small dispersive coal dust particles is in the *2* times bigger in the *2$^{nd}$* container in comparison with the *1$^{st}$* container.)

**3**. In the region from *3%* to *5%*, the dependence toward the decrease of quantity of the precipitated small dispersive coal dust slows down (the decrease to *1.5%; 3.0%; 2.6%* in the containers no. *1, 2, 3*).

**4**. Starting with the value of *5%* and up *10.5%* at the axis, $m_o/(M_o+m_o)$, the constant quantity of the coal dust particles in the containers no. *1–3* in every conducted experiment is accumulated. Herewith, the dependence of the mass share of the coal dust particles on the value of $m_o/(M_o+m_o)$ in the containers no. *1–3* has practically linear character. The mass share of the coal dust particles is increased in the *2.3* times in the container no. *1*; in the *1.8* times in the container no. *2*; in the *1.9* times in the container no. *3*, (see the graphs *1–3* in Fig. 3).

**5**. In the section no. *4*, consisting of the three containers, the percentage of the total accumulated coal dust mass in relation to the relative total introduced small dispersive coal dust mass was *17.3%*, corresponding to the following percentage distribution *3.5%; 5.1%; 8.7%* among the containers no. *1, 2, 3* accordingly. The averaged relative mass shares of the small dispersive coal dust particles in every of the three researched containers were: *(1–5.5)%; (2–8.5)%; (3–14.3)%*.

**6**. As it can be seen in the graph *4* in Fig. 2, in the region from the value of *0.5%* to *5%* at the axis, $m_o/(M_o+m_o)$, there is a clearly visible trend toward the increase of mass share of the coal dust particles, introduced to the sections no. *1–3*. Besides, the total relative mass of the dust particles, disappeared in the sections no. *1–3*, is sharply increased (on *46%*) in the region from *0.5%* to *3%*. In the region from *3%* to *5%*, the characteristic dependence increases smoothly (the increase to *9.3%*). Starting from *5%* and up to the last measured point, the approximately similar quantity of the coal dust particles fraction (*82.7%*) is jettisoned outside the model of the *IAF*.

**7**. The relative mass of the coal dust particles, accumulated in every of the three sections, was calculated as a difference between the final total mass of adsorber in the section at the completion of the experiment and the initial total mass of adsorber in the section before the experiment. The coal dust was distributed in the four sections by the following manner (from the right to the left): *(1–5.2)%; (2–7.3)%; (3–4.8)%; (4–17.3)%*. The averaged mass share of the small dispersive coal dust fraction in the four sections (from the right to the left) had the following values: *(1–2.2)%; (2–3.1)%; (3–2.2)%*.

**8**. The comparison of the total relative mass of the small dispersive coal dust particles *82.7%*, which was transported through the sections *1–3*, with the precipitated mass of the small dispersive coal dust particles (*17.3%*), which was accumulated in the sections *1–3*, makes it possible to calculate the relative mass of the small dispersive coal dust particles fraction (*65.4%*), which was jettisoned by the applied air stream outside the *IAF*.

## Influence of coal dust particles spatial distribution in granular filtering medium of adsorber on aerodynamic resistance of horizontally oriented iodine air filter

In the horizontally oriented *IAF*, it is possible to evaluate the degree of influence by the given distribution of the coal dust particles fraction in the absorber on the magnitude of the aerodynamic resistance, investigating the dependence of the aerodynamic resistance, $\Delta P$, on the volume of the air stream flow, $J$, in the case of the serially changing values (from *0%* to *10.5%*) of the relative mass share of the introduced coal dust particles, $m_o/(M_o+m_o)$, as shown in the graphs *1–15* in Fig. 4.

In the vertically oriented *IAF*, the selected curves of the dependence of the aerodynamic resistance, $\Delta P$, as a function of the air stream flow, $J$, are also shown in the graphs ***1-6*** in Fig. 4.

In the case of both the vertically oriented *IAF* as well as the horizontally oriented *IAF*, when there are no small dispersive coal dust particles fraction in the granular filtering mediums in the *IAFs*, $(m_o/(M_o+m_o)=0\%)$, the curves of the dependence of the aerodynamic resistance, $\Delta P$, as a function of the air stream flow, $J$, are completely matched as depicted in the graphs *1* and ***1*** in Fig. 4.

However, as the mass share of the introduced small dispersive coal dust particles fraction begins to increase, we found that the horizontally oriented *IAF* has the smaller magnitude of aerodynamic resistance, comparing to the magnitude of aerodynamic resistance of the vertically oriented *IAF* as shown in the graphs *15* and ***2*** in Fig 4.

In [1], it was shown that the dependence of the aerodynamic resistance, $\Delta P$, as a function of the air stream flow, $J$, can be well approximated by the following empirical expression, $\Delta P = kJ^{3/2}$, where $k$ is constant. Using this equation, we calculated the magnitude of aerodynamic resistance, $\Delta P^*$, normalized to the constant value of air stream flow, $J^* = 15\ m^3/hour$, in the selected case, when the magnitude of the *IAF*'s aerodynamic resistance was $\Delta P_{IAF} = 6000\ Pa$ as shown in Fig. 4.

In the graph *1* in Fig. 5, the measured dependence of the aerodynamic resistance, $\Delta P^*$, on the relative mass share of the introduced small dispersive coal dust particles, $m_o/(M_o+m_o)$, is shown in the case of horizontally oriented *IAF*. In the graph *2* in Fig. 5, the analogous dependence of the aerodynamic resistance on the mass share of the introduced small dispersive coal dust particles, $m_o/(M_o+m_o)$, is shown in the case of vertically oriented *IAF* ($\Delta P_{IAF} = 6000\ Pa$, $J^* = 15\ m^3/hour$) in [1].

It was interesting to make a comparison between the characteristic distributions of the small dispersive coal dust particles along the granular filtering mediums in the absorbers in the both cases of the vertically oriented *IAF* as well as the horizontally oriented *IAF*, estimating the influence by the given characteristic distribution of the small dispersive coal dust particles on the increase of the magnitude of the aerodynamic resistance in every considered case.

As it follows from the comparative analysis of data, shown in graphs *1* and *2* in Fig. 5, the dependences of the aerodynamic resistance, $\Delta P^*$, on the mass share of the introduced small dispersive coal dust particles, $m_o/(M_o+m_o)$, in the region from *0 %* to *6 %* at the axis, $m_o/(M_o+m_o)$, are identical in the cases of the vertically oriented *IAF* as well as the horizontally oriented *IAF*. This is a region, where the mass share of the small dispersive coal dust particles in all the three containers makes a direct influence on the magnitude of aerodynamic resistance in the cases of the vertically oriented *IAF* as well as the horizontally oriented *IAF*.

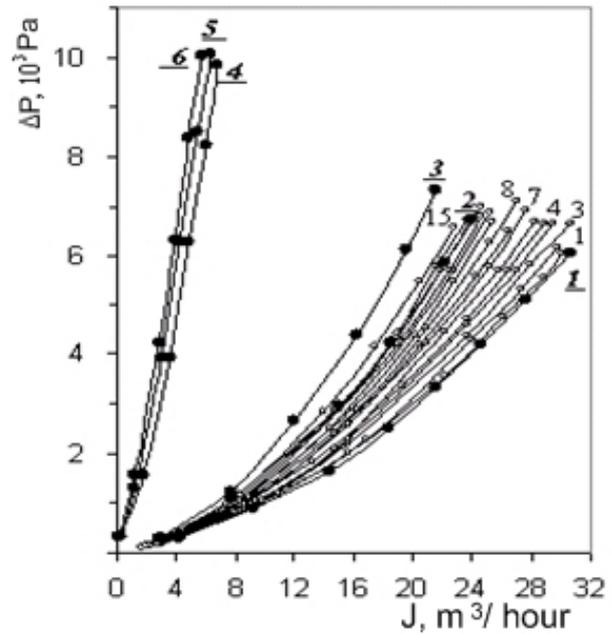

*Fig. 4. Dependence of aerodynamic resistance of absorber as function of volumetric air stream flow in cases of different values of mass share of introduced small dispersive coal dust particles in IAF, (%):*

○ – *horizontally oriented IAF:*
*1 (0), 2 (0.09), 3 (0.36), 4 (0.6), 5 (1.31), 6 (1.92), 7 (2.99), 8 (3.85), 9 (4.97), 10 (6.16), 11 (6.85), 12(7.92), 13 (9.05), 14 (9.64), 15 (10.5).*

● – *vertically oriented IAF:*
***1*** *(0),* ***2*** *(5.9),* ***3*** *(6.7),* ***4*** *(8.8),* ***5*** *(9.1),* ***6*** *(9.2).*

In the vertically oriented *IAF*, the additional introduction of the two percents of the small dispersive coal dust particles (up to *8 %* at the axis, $m_o/(M_o+m_o)$) results in the *5* times increase of the *IAF*'s aerodynamic resistance; here, it makes sense to note that the further addition of the small dispersive coal dust particles at the rate of *1.2 %* results in the exponential increase of the *IAF*'s aerodynamic resistance in the *23* times. As it is explained in [1], this increase of the *IAF*'s aerodynamic resistance provides an evidence about a prevailing influence by the sharply increased mass share of the small dispersive coal dust particles fraction (up to *24 %*) in the container no. *3* on the *IAF*'s aerodynamic resistance.

In the horizontally oriented filter, as it follows from the graph *1* in Fig. 5, the small smooth increase of the aerodynamic resistance in the *1.2* times is observed in the region from *6 %* to *10.5 %* at the axis, $m_o/(M_o+m_o)$. It is necessary to emphasis that, at the increase of mass share of the introduced small dispersive coal dust particles fraction from *0 %* до *10.5 %* at the axis, $m_o/(M_o+m_o)$, the aerodynamic resistance is increased in the *1.7* times.

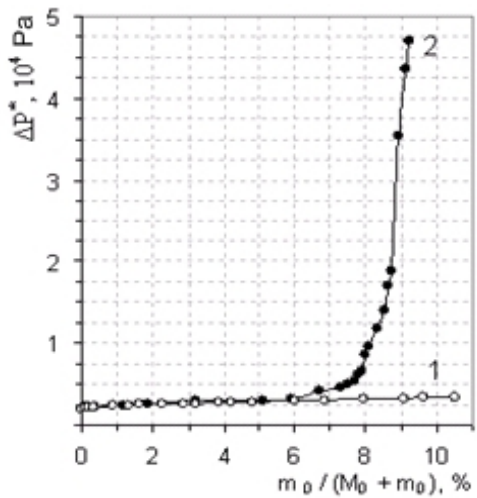

*Fig. 5. Dependence of aerodynamic resistance of adsorber, $\Delta P_{IAF} = 6000$ Pa, normalized to air stream flow, $J^* = 15$ m$^3$/hour, on mass share of introduced small dispersive coal dust particles fraction in IAF:*
*1 – horizontally oriented IAF;*
*2 – vertically oriented IAF.*

In the horizontally oriented filter, the sharp increase of mass share of the small dispersive coal dust particles is not registered in the container no. *3*, comparing to the containers no. *1* and *2*, as shown in Fig 3. As it follows from the comparative analysis of the graphs *1-3* in Fig. 3, there is an increase of the *IAF's* aerodynamic resistance due to the joint accumulation of the small dispersive coal dust particles in all the three containers, going from the precise measurements, conducted from the beginning of experiment until the completion of experiment (from *0 %* to *10.5 %* at the axis, $m_o/(M_o+m_o)$). The mass share of the accumulated small dispersive coal dust particles in any single container of the three containers has no a prevailing influence on the increase of the *IAF's* aerodynamic resistance.

After the completion of the experiment, the mass of the small disperse coal dust particles, accumulated in the segments no. *1-4* in every of the three containers, relating to the mass of the small disperse coal dust particles, precipitated in every of the three containers, was determined. As a result, the characteristic distribution of the small dispersive coal dust particles at the transverse cross-section in every of the three containers was also found. The distributions of relative mass of the small dispersive coal dust particles in the segments no. *1–4* in the containers no. *1–3* are:

|         | Seg. 1 | Seg. 2 | Seg. 3 | Seg. 4 |
|---------|--------|--------|--------|--------|
| Cont. 1 | 11.6 % | 22.0 % | 26.3 % | 40.1 % |
| Cont. 2 | 6.2 %  | 20.7 % | 38.2 % | 34.9 % |
| Cont. 3 | 10.9 % | 24.2 % | 39.6 % | 25.3 % |

The graphical representation of averaged research results is shown in Fig. 6.

As it follows from the given experimental data, the smallest quantity of the small disperse coal dust particles is concentrated in the close proximity to the absorber's input surface, and its mass share is gradually increased, reaching the maximum in the segment *no. 3* in the containers №2, 3, and it's mass share is sharply decreased in the segment *no. 4* in the containers № 2, 3 in the filter. The relative mass share of small disperse coal dust is almost linearly increased along the transversal cross-section in all the segments, reaching its maximum in the bottom segment *no. 4* in the container №1 in the IAF.

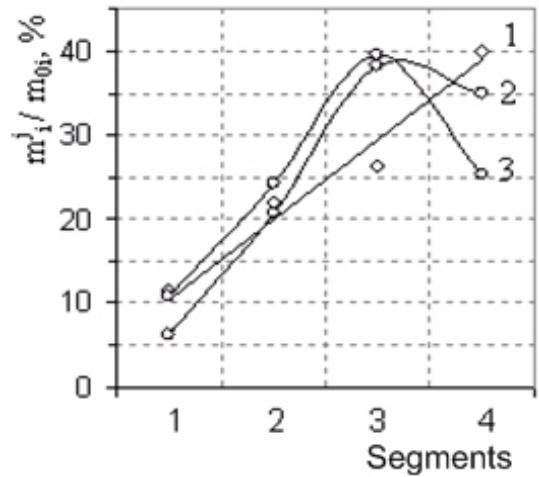

*Fig. 6. Distribution of relative mass share of introduced small dispersive coal dust particles fraction in transverse cross-section of containers (№1-3):*
*1 – №1; 2 – №2; 3 – №3*
*( i - от 1 до 3; j – от 1 до 4).*

In Fig. 7, the picture of the upper side surface of the section *no. 2* in the absorber in the *IAF* is shown. It is necessary to emphasis that the small disperse coal dust layer is absent in the granular filtering medium with the cylindrical coal granules in the filter during the model operation (the photography is made at the top of the *IAF*).

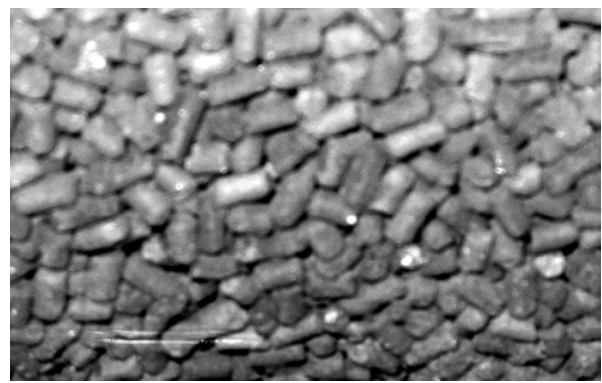

*Fig. 7. Photography of upper side surface of part of 2$^{nd}$ section in absorber in IAF (see Fig. 1 (a)).*
*Small disperse coal dust layer is absent in granular filtering medium with cylindrical coal granules in absorber during IAF's model operation.*
*(Photography is made at top of IAF).*

Applying the visual observation to the transversal cross-section of sections no. *1–3*, it was determined that the quantity of the deposited small disperse coal dust particles is increased from the top to the bottom, reaching the quantitative concentrative maximum at the bottom as shown in the photographs in Figs. 8, 9.

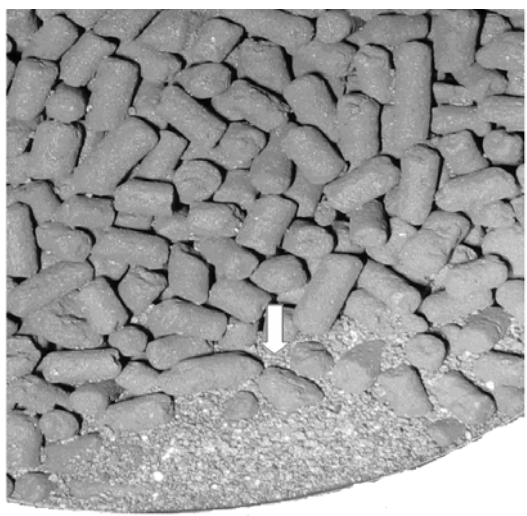

**Base of Iodine Air Filter (IAF)**

*Fig. 8. Photography of cross-section of part of 3<sup>rd</sup> section in absorber in Iodine Air Filter (IAF) (see Fig. 1 (a)). Arrow shows appearance of small disperse coal dust layer in granular filtering medium with cylindrical coal granules in absorber during IAF's model operation.*

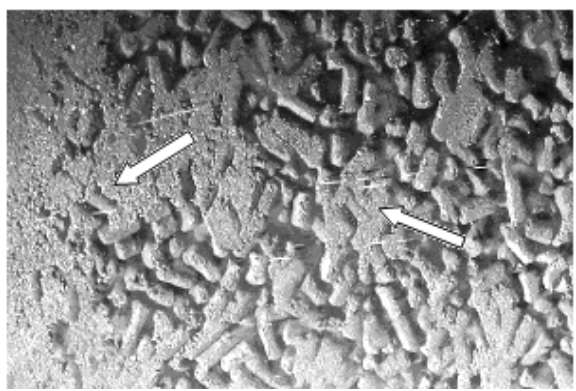

*Fig. 9. Photograph of bottom side surface of section no. 2 in absorber in Iodine Air Filter (IAF) (see Fig. 1(a)). Arrows point out to appeared layer of accumulated small dispersive coal dust particles in granular filtering medium with cylindrical coal granules in absorber during IAF's model operation.*

As shown in Fig. 10, the mass share of the small disperse coal dust particles in the researched four segments in the containers no. *1-3* in the model of the *IAF* has the following values:

|          | Seg. 1 | Seg. 2 | Seg. 3 | Seg. 4 |
|----------|--------|--------|--------|--------|
| Cont. 1  | 2.5 %  | 4.7 %  | 5.4 %  | 9.2 %  |
| Cont. 2  | 2.8 %  | 6.7 %  | 11.5 % | 10.7 % |
| Cont. 3  | 8.1 %  | 14.7 % | 17.7 % | 14.2 % |

As it follows from the obtained experimental data, the mass share of the small disperse coal dust in all the four segments (except for the second segment in the container no. *3*) does not reach the critical value of *18 %* at which the small disperse coal dust particles layer can block the air stream flow in the filter. Thus, the layer of the granular filtering medium, containing the cylindrical coal granules with the accumulated small disperse coal dust particles fraction, continues to be well transparent for the small disperse coal dust particles movement in the subsequent stages of the *IAF's* operation during the experimental measurements.

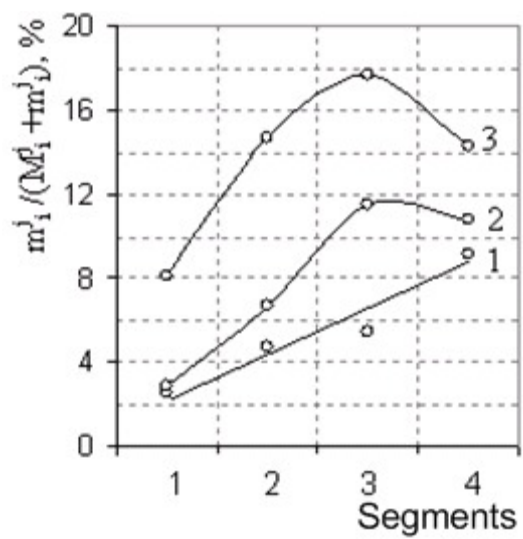

*Fig. 10. Dependence of distribution of mass share of accumulated small dispersive coal dust particles fraction in segments no. 1-4 across transverse cross-section of containers no. 1-3 in absorber, filled with granular filtering medium with cylindrical coal granules, during IAF's model operation:*
*1 - №1; 2-№2; 3-№3; (i - from 1 to 3; j – from 1 to 4).*

In the vertically oriented filter, when the mass share of the small dispersive coal dust particles reaches *9.1 %*, the magnitude of aerodynamic resistance is in the *14.4* times bigger, comparing to the magnitude of aerodynamic resistance in the horizontally oriented *IAF* in the case with the same mass share of the small dispersive coal dust particles.

In the horizontally oriented *IAF*, the introduction of the mass share of the small dispersive coal dust particles up to *10.5 %* results in an increase of its aerodynamic resistance in *1.7* times.

## Conclusion

Going from the completed comparative analysis of obtained research results on the small disperse coal dust masses distribution in both the *IAF* with the vertical orientation and the *IAF* with the horizontal orientation, the following conclusions can be made:

**1**. The physical character of the small disperse coal dust distribution in the *IAF* with the horizontal orientation significantly differs from the physical character of the small disperse coal dust distribution in the filter with the vertical orientation during the small disperse coal dust particles fracture appearance in the sub-surface layer of absorber's input as a direct result of the air-dust aerosol stream flow during the *IAF's* operation.

In the vertically filter the small disperse coal dust fracture is mainly accumulated in the thin sub-surface layer of absorber in the *IAF*. At the end of experiment, the mass share of the small disperse coal dust particles in the upper container no. *3* reaches the value of *24%* at $m_o/(M_o+m_o)=9.1\%$, resulting in the *23* times increase of the *IAF's* aerodynamic resistance.

In the horizontally oriented filter, the dependence of the mass share of the introduced small disperse coal dust particles is practically linear in the region, beginning with the value of *5%* at the axis, $m_o/(M_o+m_o)$, in the containers no. *1–3*. There is no sharp increase of mass share of the small dispersive coal dust particles in the container no. *3*, which is situated close to the source of dust. The *IAF's* aerodynamic resistance, only increases in *1.7* times at the end of experiment ($m_o/(M_o+m_o)=10.5\%$).

In the vertically oriented filter, the level of mass share of the small dispersive coal dust particles, accumulated in the container no. *3*, makes a main impact on the magnitude of the *IAF's* aerodynamic resistance.

In the horizontally filter, the mass share of the small dispersive coal dust particles, accumulated in all the three containers, has a joint influence on the increase of magnitude of the *IAF's* aerodynamic resistance.

**2**. In the vertically oriented filter, the mass share of the small dispersive coal dust particles, accumulated in the containers, is homogenously distributed at the transverse cross-section of every container. That is why, the small dispersive coal dust particles accumulation in the sub-surface thin layer in the granular filtering medium in the absorber results in the *IAF's* malfunction. At the concentration of the small dispersive coal dust particles equal to the critical concentration of *18 %* at the axis, $m_o/(M_o+m_o)$, this sub-surface layer in the absorber's input becomes a practically impenetrable layer for the air stream flow, resulting in the *IAF's* complete operational failure.

In the horizontally oriented filter, the mass share of the small dispersive coal dust particles, accumulated in the containers no. *1–3*, is distributed non-homogeneously at the transverse cross-section of every container. The mass share of the small dispersive coal dust particles, accumulated in the containers no. *1-3*, has a minimum at the segment no. *1*, then it increases, reaching its maximum in the bottom segment no. *4* in the container no. *1* and in the segment no. *3* in the containers no. *2* and *3* in the model of the *IAF*. The mass share of the small dispersive coal dust particles does not reach its critical value in all the segments in the containers no. *1-3*. Thus, in every of the three containers, the granular filtering medium, consisting of the mixture between the cylindrical coal granules and the precipitated small dispersive coal dust particles, does not create an obstacle for the air stream flow in the *IAF's* model at the subsequent stages of its operation.

3. In the horizontally oriented filter, the introduction of the small disperse coal dust masses at the rate of *1.5%* from the full mass of absorbent in the *4th* container appears to have no any negative influence on the operation of absorber in distinction from the case of the *IAF* with the vertical orientation of adsorber.

In the horizontally oriented filter, the small disperse coal dust masses are transported to the lower part of the absorber, making the upper part of its cross-section free of the particles with the big dimensions, which can originate the coal dust structures creation, increasing the aerodynamic resistance and blocking the normal air flow.

———————